\documentclass[prl,twocolumn,amsmath,amssymb,showpacs]{revtex4}

\usepackage{graphicx}
\usepackage{dcolumn}
\usepackage{bm}
\usepackage{color}

\begin{document}

\preprint{preprint}

\title{Field-Induced Transition on Triangular Plane in the Spin Ice Compound Dy$_2$Ti$_2$O$_7$}

\author{Ryuji Higashinaka}
\affiliation{
Department of Physics, Kyoto University, Kyoto 606-8502, Japan
}

\author{Yoshiteru Maeno}
\affiliation{
Department of Physics, Kyoto University, Kyoto 606-8502, Japan
}
\affiliation{
International Innovation Center, Kyoto University, Kyoto 606-8501, Japan
}

\date{\today}

\begin{abstract}
 The origin of the lowest-temperature anomaly reported several years ago using a polycrystalline sample of the spin ice compound Dy$_2$Ti$_2$O$_7$ had remained unresolved. Here we finally clarify its origin by susceptibility measurements down to 65 mK using single crystals under accurate control of the magnetic fields in two independent directions. We demonstrate that the transition is induced under a subtle field combination that precisely cancel the nearest-neighbor spin interactions acting on the spins on the triangular lattice within the pyrochlore structure. Contrary to the other two field-induced transitions, this transition is driven only by the interactions beyond the nearest neighbors. Our observation thus provides the first qualitative evidence for the essential importance of the dipolar interaction beyond the nearest neighbors in the spin ice.
\end{abstract}

\maketitle

 Recently, the physics of geometrically frustrated systems has attracted a great deal of attention, because a variety of unconventional spin behavior emerges as a result of the suppression of ordinary long-range ordered states.
Among them, the spin ice behavior is peculiar since the spin frustration originates from ferromagnetic (FM) interactions, in contrast with frustration due to antiferromagnetic (AFM) interactions commonly found in systems based on triangular spin configuration.
In spin ice compounds such as Ho$_2$Ti$_2$O$_7$ and Dy$_2$Ti$_2$O$_7$~\cite{Harris97,RamirezN}, crystal field effects endow the ground state of Ho$^{3+}$ and Dy$^{3+}$ ions with a doublet with strong Ising anisotropy along the local $\langle 111 \rangle$ direction~\cite{JanaDy2}.
Since these ions have large magnetic moments, the dominant interaction is FM dipolar interaction $D_{{\rm nn}}$, not AFM exchange interaction $J_{{\rm nn}}$, and the effective nearest neighbor (nn) interaction $J_{{\rm eff}}$ (= $D_{{\rm nn}} + J_{{\rm nn}}$) is FM.
Throughout this paper, we use the term ``spin'' to designate the magnetic moment of a rare earth ion.
The FM-nn interaction leads to the ``2-in 2-out'' ground state in which two of the four spins on each tetrahedron point inward and the other two outward.
Such a ``2-in 2-out" constraint, called the ice rule from the analogy to the proton ordering in water ice, and the resulting spin configurations induce a macroscopically degenerate ground state with a residual entropy.\\
\quad At first, the spin ice behavior was expressed by the near neighbor spin-ice model considering only $J_{{\rm eff}}$~\cite{Harris97}.
However, in order to explain some key experimental features quantitatively the dipolar spin-ice model was introduced, in which the importance of long-range dipolar interaction is emphasized~\cite{Hertog00}.
By Monte Carlo simulations based on this model, it was possible to reproduce the results of powder neutron diffraction patterns~\cite{Bramwell01}.
The model further predicts a phase transition to the true ground state using a spin loop move algorithm~\cite{Melko01}, although the predicted transition has not been observed experimentally down to 65~mK~\cite{Fukazawa02}.
In addition, it has recently been shown that long-range dipolar interactions are largely screened but the remaining medium-range interaction within several neighbors plays an important role in the $q$-dependent susceptibility and is the cause of the ordering~\cite{Isakov05}.
For definitive evidence of the importance of the dipolar interaction apart from the nn interaction, it is crucial to demonstrate a phenomenon induced directly by the further neighbor interactions.\\
\quad An outstanding problem related to the above discussion concerns the origin of the peak in the specific heat divided by temperature $C/T$ around 0.35~K observed in polycrystalline Dy$_2$Ti$_2$O$_7$ by Ramirez $et$ $al$.~\cite{RamirezN}. 
They reported three peaks at 1.2, 0.45 and 0.35~K in $C/T$.
Curiously, the peak temperatures are almost independent of the applied magnetic field strength.
From the specific heat for the [110] field direction~\cite{Hiroi03110}, the origin of the peak at 1.2~K has been ascribed to the ordering of spin chains with the Ising axes {\it perpendicular} to the field direction~\cite{Yoshida04,Ruff04}.
In contrast, the origin of the 0.45~K peak is most probably related to the transition from ``Kagome-ice state'' to 1-in 3-out ordered state induced when magnetic field is {\it parallel} to the [111] direction~\cite{Hiroi03,Higashin03,Sakakibara03,Higashin04}.
However, to the best of our knowledge there have been no experimental reports clarifying the origin of the lowest temperature peak at 0.35~K.\\
%
%
\begin{figure}[btp]
\includegraphics[width=0.8\linewidth]{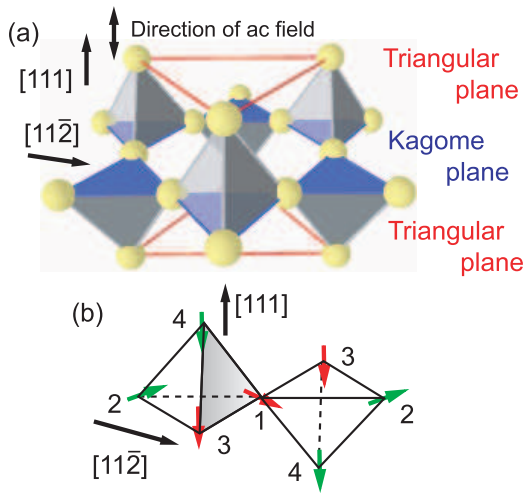}
\caption{\label{FieldDirection} (color online) (a) The relation between the pyrochlore lattice and two field directions.
A pyrochlore lattice consists of an alternating stacking of the triangular (red) and Kagome planes (blue) along the [111] direction.
The [11$\bar{2}$] direction is in the Kagome plane.
(b) One of the possible spin configurations of a pair of tetrahedra.
The arrows at the vertices of a tetrahedron represent the magnetic moment of Dy$^{3+}$ ions.}
\end{figure}
\quad Very recently, J. Ruff $et$ $al$. theoretically proposed an origin for the 0.35~K peak \cite{Ruff04}.
The pyrochlore lattice can be viewed as an alternating stacking of the triangular and Kagome planes normal to the [111] direction (Fig.~1(a)).
In a range of field direction within the Kagome plane centered at the [11$\bar{2}$] direction the directions of all the spins on the Kagome planes are determined by this field. 
However, the spins on the triangular planes are perpendicular to this field and are experiencing no Zeeman interaction.
Thus, the direction of these spins is decided only by the effective nn interaction, $J_{{\rm eff}}$.
As shown in Fig.~1~(b), three spins (1, 2 and 3) on the Kagome plane are fixed in a 1-in 2-out configuration for the up-tetrahedron and 2-in 1-out for the down-tetrahedron.
Then the ice-rule constraint makes {\it all} the spins on the triangular planes (4) point along the [$\bar{1}\bar{1}\bar{1}$] direction.
In this situation, an additional [111] magnetic field may cancel out $J_{{\rm eff}}$ on spins labeled (4).
These spins (4) are among themselves third nearest neighbors on the pyrochlore lattice.
Combined spins on different triangular layers form a face centered cubic (FCC) lattice when effects from the spins on the Kagome planes are neglected.
Under such a canceling field, the residual spin-spin interactions, which mainly consists of long-range dipolar interaction and the third nearest-neighbor exchange interaction $J_3$, become important.
Such interactions are expected to induce a FM ordering of these spins if $J_3$ is not too strongly AFM~\cite{Ruff04}.
Here we experimentally identify the predicted field-induced transition as the origin of the anomaly at low temperature.\\
%
\quad We used single crystals of Dy$_2$Ti$_2$O$_7$ grown by a floating zone method~\cite{Fukazawa02}, and subsequently annealed at 700~$^{\circ}$C in O$_2$ for 24~hours.
Two samples were cut from single crystalline rods into a thin plate with the plane normal to the [111] direction and a pencil-shape rod parallel to the [111] direction.
The size and mass of the two samples were approximately 1.7~$\times$ 1.7~$\times$ 0.7~mm$^3$ and 14~mg, and 0.8~$\times$ 1.0~$\times$ 3.2~mm$^3$ and 19~mg, respectively. 
The [11$\bar{2}$] direction lies on the (111) plane.
The real component $\chi '$ and the imaginary component $\chi ''$ of ac susceptibility were measured by a mutual inductance method down to 65 mK using a commercial dilution refrigerator.
The strength of the ac field was 0.05~Oe-rms along the [111] direction.
In order to control the magnetic field along the [11$\bar{2}$] and [111] directions independently, we used a ``vector magnet'' system consisting of two superconducting magnets~\cite{Deguchi04VecMag}.
One of the magnets points along the horizontal axis $H_x$ and the other one along the vertical axis $H_z$ (aligned to the [111] direction).
We can apply fields up to 5~T for $H_{x}$ and 3~T for $H_{z}$.
These two magnets are rotated around the cryostat using a pulse motor with an angular precision of 0.01~degree.
To ensure good thermal contact, we evaporated gold films on the sample surfaces and linked the sample to the heat bath using a 100-$\mu$m gold wire.
We used a pair of Hall sensors for obtaining an accurate field alignment along the [11$\bar{2}$] and [111] directions.
We determine the azimuthal angle $\phi_{{\rm K}}$ in the Kagome plane between the directions of $H_x$ and [11$\bar{2}$] by the signal of Hall sensors; at $\phi_{{\rm K}}$ = 0, $H_x$ is parallel to the [11$\bar{2}$] direction.
The overall accuracy of the field alignment including the resolution of the Hall sensors and misalignment between the Hall sensors and the crystal axes were within 2 degrees for the [11$\bar{2}$] direction and 1 degree for the [111] direction.
Because the field direction changes continuously, it is difficult to estimate an accurate demagnetization factor, and we present data here with no demagnetization correction.\\
%
%
\begin{figure}[btp]
\includegraphics[width=0.7\linewidth]{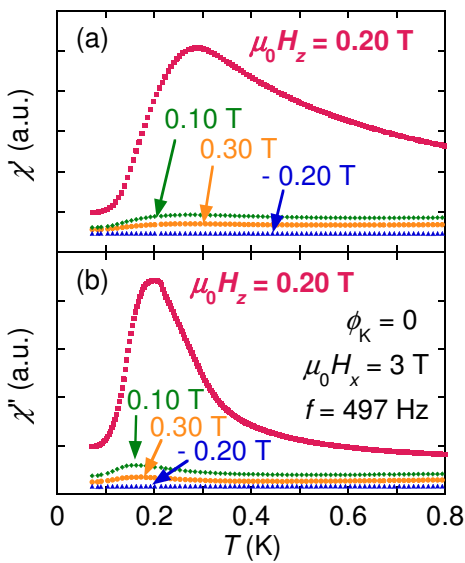}
\caption{\label{Tdep4T} (color online) Temperature dependence of real and imaginary parts of the ac susceptibility (a) $\chi '$ and (b) $\chi ''$ of Dy$_2$Ti$_2$O$_7$.
Note that both components are highly asymmetric in $H_z$.
Features below 0.3 K for $\mu_0 H_z$ = 0.20 are attributable to a magnetic transition.}
\end{figure}
\quad In the measurements, we first applied $\mu_{0} H_x$ = 3~T along the [11$\bar{2}$] direction in order to hold the spins on the Kagome planes and then varied an additional $H_z$ along the [111] direction.
At $\mu_{0} H_x$ = 3~T, the Zeeman energy for the spins 1 (Fig.~1~(b)) is $E_{{\rm Z}}^{(1)} = g_{J}J \mu_{{\rm B}} (\mu_{0}H_x) {\rm cos}\theta_{1}$ = 19.1 K and for the spins 2 and 3, $E_{{\rm Z}}^{(2)} = E_{{\rm Z}}^{(3)} = g_{J}J \mu_{{\rm B}}(\mu_{0}H_x) {\rm cos}\theta_{2}$ = 9.6~K.
Here $g_{J}$ is Lande's g factor, $J =~15/2$ is the total angular momentum, $\mu_{{\rm B}}$ is the Bohr magneton, $\mu_{0}$ is permittivity and $\theta_{i}$ is the angle between the field direction and Ising axis at site $i$.
These values are much greater than both the effective nn interaction, $J_{{\rm eff}}$ = 1.11 K~\cite{Hertog00} and the measurement temperatures in this experiment.
Therefore, at relatively small $H_z$/$H_x$, all the spins on the Kagome planes are fixed by $H_x$.\\
\quad In Fig.~2, we show the temperature dependence of $\chi '$ and $\chi ''$ at $\mu_{0} H_{x}$ = 3 T, at various $\mu_{0} H_{z}$ and at 497 Hz for $\phi_{{\rm K}}$ = 0$^{\circ}$ for a pencil-shape sample.
Both components are highly asymmetric with respect to $H_z$.
At $\mu_{0}H_{z}$ = 0.20 T, the changes are most pronounced.
In contrast, at $\mu_0 H_z$ = - 0.20 T and at other values of $H_z$ both components change little with temperature.
Such enhancements around 0.2 T are also observed clearly in $H_z$ dependence discussed later (Fig.~3).
In addition, even at higher temperature ($\sim$ 0.8 K), both components are enhanced at $\mu_0 H_z$ = 0.2 T.
This large response to the small ac field strongly suggests the spins on the triangular planes are freed from the spin-ice constraint and the external fields, and become paramagnetic.
This is exactly expected by the theory of Ruff $et$ $al$.~\cite{Ruff04} that the small [111] field ($\sim$ 0.24 T) cancel out $J_{{\rm eff}}$ for spins on the triangular planes.\\
\quad At $\mu_0 H_z$ = 0.20 T, there is a broad peak centered at 0.29 K in $\chi'$ and a peak at 0.20 K in $\chi''$.
We also observed these peaks at almost the same temperature $T$~=~0.28~$\pm$~0.01~K for different frequencies (50, 1.75 k and 3 kHz; data not shown).
This invariance of the peak temperature indicates that the peaks are associated with a transition to long-range magnetic order rather than with spin freezing~\cite{CommentPRLAcChi}.
The results of thin plate sample are qualitatively the same as those of pencil-shape sample (data not shown).\\
\quad In Fig.~3 we show the $H_z$ dependence of $\chi '$ at $\mu_{0} H_x$ = 4 T and at 0.30 K for $\phi_{{\rm K}}$ = 0$^{\circ}$ and 180$^{\circ}$.
Most importantly, we find that at $\phi_{{\rm K}}$ = 60 $\times 2n$ deg. ($n$ = integer), $\chi '$ exhibits a peak at a certain positive $H_z$ ($\mu_0 H_z$ $\sim$ 0.2 T, $\phi_{{\rm K}}$ = 0$^{\circ}$).
In contrast, at $\phi_{{\rm K}}$ = 60($2n + 1$) deg. the peak in $\chi '$ appears at the corresponding negative $H_z$ ($\mu_0 H_z$ $\sim$ - 0.2 T, $\phi_{{\rm K}}$ = 180$^{\circ}$).
This field reversal is indeed expected by the difference of the stable spin configuration under different $H_x$ direction.
When we rotate $H_x$ by 60 degrees, the stable spin configuration on the Kagome lattice changes from 2-in 1-out to 1-in 2-out for the up-tetrahedron and vice versa for the down-tetrahedron (Fig.~1(b)).
At the same time, the stable spin direction on the triangular planes changes from the [$\bar{1}\bar{1}\bar{1}$] to [111] direction and the sign of the required canceling field $H_z$ changes.
The observed asymmetry in $H_z$ and the sign reversal of the characteristic $H_z$ by the rotation of $H_x$ strongly suggest that this anomaly is indeed due to the predicted spin ordering on the triangular planes~\cite{Ruff04}.\\
\quad If the field alignment is accurate, the absolute value of the peak field $H_z$ should be the same.
In reality, there is a little misalignment of both fields in the experiment.
A small difference among the peak fields at different $\phi_{{\rm K}}$ is attributable to a small misalignment of $H_x$ from the (111) plane.
From the $H_x$ dependence of the peak field (data not shown), we can evaluate the critical field by extrapolating the peak field in $H_z$ for various $H_x$ to $\mu_0 H_x$ = 0 T and the misalignment angle by considering the difference between the critical field and actual peak value.
From this procedure using the data at $n$ = 0 to 5 for $\phi_{{\rm K}}$, we evaluated the critical field $\mu_0 H_c$ as 0.25 $\pm$ 0.01 T and the misalignment as about 2 degrees.
This $H_c$ is consistent with the theoretical expectation. 
The expected Zeeman energy of a spin on the triangular planes by the combined nn exchange and dipolar energies is 1.71 K considering long-range dipolar interaction by the Ewald method \cite{Ruff04}.
Thus, the scale of the critical field $H_{{\rm c}}$ is estimated as 1.71 K/($g_{\rm J}\langle J_z \rangle\mu_{{\rm B}}/k_{{\rm B}}$) = 0.25 T.\\
%
%
\begin{figure}[btp]
\includegraphics[width=0.6\linewidth]{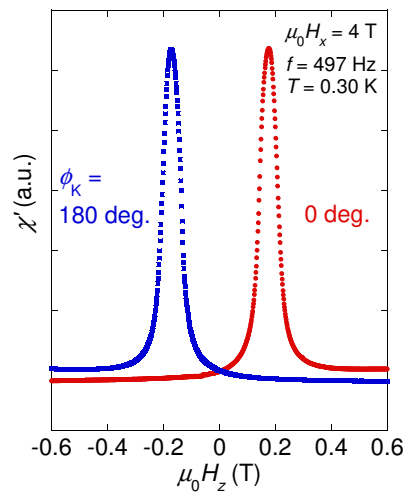}
\caption{\label{PhaiDep} (color online) Dependence of the real part of the ac susceptibility $\chi'$ on the field strength along the [111] direction for $\phi_{{\rm K}}$ = 0$^{\circ}$ and 180$^{\circ}$.
When $H_x$ is rotated by 180 (60 $\times$ 3) degrees, the sign of the peak field changes.}
\end{figure}
%
%
\begin{figure}[btp]
\includegraphics[width=0.7\linewidth]{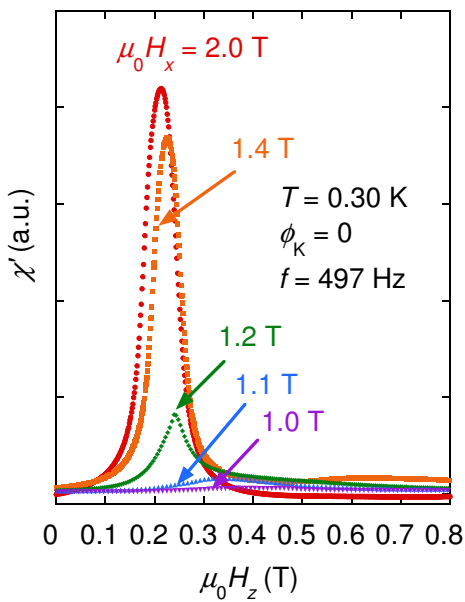}
\caption{\label{HzDep280mK} (color online) Variation of the ac susceptibility peak with the field $H_{z}$ along the [111] direction and $H_{x}$ along the [11$\bar{2}$].
The peak around $\mu_{0} H_z$ = 0.22 T disappears below $\mu_{0} H_x$ = 1.2 T.}
\end{figure}
\quad In Fig.~4, we show the $H_z$ dependence of $\chi '$ and $\chi ''$ at 0.3 K and at various $H_x$ for $\phi_{{\rm K}}$ = 0$^{\circ}$.
With decreasing horizontal field $H_x$, the peak around $\mu_{0} H_z$ = 0.22 T disappears below $\mu_{0} H_x$ = 1.2 T.
This is compatible with the behavior of $C/T$ of polycrystals reported by Ramirez $et$ $al$.~\cite{RamirezN} for which the peak appears only for fields above 1 T.
This provides additional evidence that the origin of the low temperature peak in $C/T$ is the ordering of the spins on the triangular planes.\\
\quad These anomalies in $\chi'$ and $\chi''$ emerge under the condition that the effective nn interaction for the triangular spins are canceled out and the long-range dipolar and further neighbor exchange interaction between these spins are important.
Until now, there have been some discussions about the importance of the long-range interactions for explaining the physics of spin ice, for example, fitting the specific heat $C_{{\rm v}}$~\cite{Hertog00,Bramwell01} and the neutron structure factor $S({\bm q})$~\cite{Bramwell01}.
In these discussions while of quantitative relevance, the importance of the long-range interaction was not made so qualitatively clear and important.
Our report here marks the first experimental demonstration of a phenomenon in the spin ice originating from the long-range interaction. \\
\quad The observed critical field is in good agreement with the theoretical expectation.
The predicted process also explains the behavior in polycrystals: grains satisfying the critical condition $\mu_0 H_{[111]}$ = $\mu_0 H_{{\rm c}}$ = 0.25 T with $\mu_0 H_{[11\bar{2}]}~ \textgreater$ 1.2 T exhibit magnetic transitions one after another with increasing magnetic field.
In the Monte Carlo simulation~\cite{Ruff04}, using a simple dipolar spin ice model with long-range dipolar interaction and only the nn exchange interaction $J_{\rm nn}$, FM ordering of the spins on the triangular planes occurs at 0.65 K.
When the AFM third nearest-neighbor exchange interaction $J_3$ = - 0.027 K is included ($J_3$ = 0 in the simple dipolar spin ice model), the transition shifts to 0.28 K with the ordering of the spins on the triangular planes still FM.
Although it is difficult to conclude solely from the present ac susceptibility measurement that this ordering is FM ordering as theoretically predicted, the observed large peaks in the dissipation $\chi''$ are consistent with the motions of FM domains. \\

\quad The estimated value of $J_3$ from this procedure, however, is not consistent with that to reproduce the transition under the [110] field, which is well fitted with $J_3$ = 0.
To resolve this discrepancy, one may have to reconsider the exchange parameters beyond nearest neighbors in the dipolar spin ice model~\cite{Hertog00}, including second neighbor exchange interaction $J_2$. 
For the field-induced transition reported here, the role of $J_2$ is merely to renormalize the value of $H_c$ slightly.\\
\quad In summary, by ac susceptibility measurements of single crystals of the spin ice Dy$_2$Ti$_2$O$_7$ under well-controlled magnetic fields, we finally clarified the origin of the unresolved anomaly reported in the polycrystals.
We clarified its origin as a field-induced magnetic transition governed by the long-range dipolar interaction beyond the nearest neighbors, as predicted by the recent theory by Ruff $et$ $al$. \\
\quad The authors are grateful to M. Gingras, J. Ruff and R. Melko for useful discussion and sending their results on the theoretical expectation before publication, to K. Deguchi for his technical help as well as useful advices and to H. Fukazawa and S. Nakatsuji for useful discussion.
This work has been supported by Grants-in-Aid for Scientific Research (S) from JSPS and for the 21 Century COE ``Center for Diversity and Universality in Physics'' from MEXT of Japan.


\end{document}